\documentclass[preprint,prc,a4paper,epsf,nofootinbib]{revtex4-1}
\usepackage{epsfig}
\usepackage{makecell}
\usepackage{array}
\long\def\symbolfootnote[#1]#2{\begingroup%
\def\thefootnote{\fnsymbol{footnote}}\footnote[#1]{#2}\endgroup}
\newcommand{\boldm}[1]{\mbox{\boldmath{$#1$}}}
\newcommand{\Del}{$\Delta(1232)$ }
\def\CA{{\cal A}} \def\CB{{\cal B}} \def\CC{{\cal C}}
\def\CD{{\cal D}}  
  \def\CI{{\cal I}}
  \def\CL{{\cal L}}
\def\CM{{\cal M}}  
  
\def\CS{{\cal S}} \def\CT{{\cal T}}

\def\prc#1{Phys.\ Rev.\ C\ {\bf #1}}

\def\npa#1{Nucl.\ Phys.\ A\ {\bf #1}}

\def\epja#1{Eur.\ Phys.\ J.\ A {\bf #1}}

\def\be{\begin{equation}}
\def\ee{\end{equation}}
\def\Be{\begin{eqnarray}}
\def\Ee{\end{eqnarray}}
\def\ba{\begin{array}}
\def\ea{\end{array}}

\newcommand{\PreserveBackslash}[1]{\let\temp=\\#1\let\\=\temp}

\begin{document}
\title{Chiral perturbation theory calculation for $p n \to d \pi \pi$ at threshold}
\author{B. Liu$^{1,2}$, V. Baru$^{2,3}$, J. Haidenbauer$^{2,4}$ and C. Hanhart$^{2,4}$}

\affiliation{$^1$School of Science, Xian Jiaotong University, Xian 710049,
China  \\
$^2$Institut f\"{u}r Kernphysik and J\"ulich Center
            for Hadron Physics, Forschungszentrum J\"{u}lich,
            D--52425 J\"{u}lich, Germany\\
$^3$Institute for Theoretical and Experimental Physics,
 117218, B. Cheremushkinskaya 25, Moscow, Russia\\
$^4$Institute for Advanced Simulation,
            Forschungszentrum J\"{u}lich, D--52425 J\"{u}lich, Germany }

\date{}
\begin{abstract}
We investigate the reaction $pn\rightarrow d \pi\pi$
in the framework of Chiral Perturbation Theory.
For the first time a complete calculation of the leading order
contributions is presented.
We identify various diagrams 
that are of equal importance as compared to
 those recognized in earlier works. The diagrams  at leading order behave
  as expected by the power counting.
Also for the first time the nucleon--nucleon interaction in the
initial, intermediate and final state is included consistently
and found to be very important.  This study provides a theoretical basis
for a controlled evaluation of the non--resonant contributions in two--pion production
reactions in nucleon--nucleon collisions.

\end{abstract}
\maketitle

\section{Introduction}

As the first strong inelastic threshold of the nucleon--nucleon system, pion
production in $NN$ collisions has attracted a large number of theoretical as
well as experimental works. Due to advances in the experimental methods
measurements in the threshold region became possible in the beginning of the
1990ies, and it quickly became clear that the theoretical models existing at the
time~\cite{koltunundreitan} were not able to describe these data --- see
Ref.~\cite{report} and references therein. They fell short by a factor of
two for reactions with an isoscalar $NN$ pair in the final state, while there
was a discrepancy of even more than an order of magnitude for those with an
isovector $NN$ pair in the final state. The numerous efforts to improve the
phenomenological approaches~\cite{heavymeson,offshell}, although quite
successful for various observables~\cite{ourpols}, did lack for systematics,
gave contradictory answers, and involved a sensitivity to unobservable short range effects.

To overcome those deficiencies, in recent years considerable theoretical efforts
went into the development of an effective field theory that can be applied to the
reactions $NN\to NN\pi$. In early studies it became clear, however, that the
original power counting by Weinberg~\cite{Weinberg:1990rz,Weinberg:1991um}
needs modifications in order to arrive at a convergent
expansion~\cite{NNpi,NNpicharged} (see also Ref.~\cite{BKMnovel} where it was pointed
out that the naive power counting using the heavy baryon formalism is not applicable above the
pion production threshold --- the necessary adaptions are outlined in
Ref.~\cite{withandreas}).
Indeed, for
neutral pion production in $pp$ collisions, the corrections due to
the next-to-leading order (NLO) increased the discrepancy with the data and,
moreover, the next-to-next-to-leading order (NNLO) contributions turned out to
be even larger than the NLO terms ~\cite{NNpiloops}.
The origin of these difficulties was identified quite early by Cohen et
al.~\cite{bira1}, see also ~\cite{rocha}, who stressed that the
additional new scale, inherent in reactions of the type of
$NN\to NN\pi$, needs to be accounted for in the power counting.
Since the two nucleons in the initial state need to have sufficiently
high kinetic energy to produce the on-shell pion in the final state, the
initial center-of-mass momentum needs to be larger than
\begin{equation}
p_{\rm thr}^{(1)} = \sqrt{M_N \, m_\pi}\,,
\quad
\mbox{with}
\quad
\frac{p_{\rm thr}^{(1)}}{\Lambda_\chi} \simeq  \ 0.4 \,,
\label{expand}
\end{equation}
where $m_\pi$ and $M_N$ refer to the pion and nucleon mass, respectively,
and $\Lambda_\chi$ denotes the typical hadronic scale, here identified with
$M_N$ to get a numerical estimate for the expansion parameter.
The proper way to include this scale was presented in Ref.~\cite{ch3body}
and implemented in Ref.~\cite{withnorbert}, see Ref.~\cite{report} for a
review article. As a result $p$-wave pion production is governed by
tree-level diagrams up to NNLO \cite{ch3body,p-wave} in the modified power counting
scheme. On the other hand, for $s$-wave pion production,
pion loops, studied in detail in Ref.~\cite{lensky1}, start to contribute already at NLO.
The individual loop contributions turned out to grow linearly with increasing $NN$
relative momentum which resulted in a large sensitivity to the employed $NN$ wave
functions ~\cite{Gardestig}. However, it was demonstrated in Refs.~\cite{lensky1,Baru_proc} that all irreducible
loop contributions at NLO  cancel altogether
which was needed to  keep the scheme self-consistent.
Furthermore, the proper separation of irreducible terms at NLO from reducible
ones in the loops 
resulted in an increase of  the
most important LO operator for charged pion production, first investigated in
Ref.~\cite{koltunundreitan}, by a factor of 4/3. This increase  was sufficient to overcome the
apparent discrepancy with the data in the reaction $pp\to d\pi^+$.
The neutral pion channel is more challenging --- it still calls for a calculation
of subleading loop contributions. First steps in this direction
were taken in Refs.~\cite{withandreas,kim}.
We further emphasize that the $\Delta$(1232) isobar should be
taken into account explicitly as a dynamical degree of freedom~\cite{bira1}
because the Delta-nucleon mass difference, $\Delta M$, is also of the order of $p_{\rm thr}$.
This general argument was confirmed numerically in phenomenological
calculations~\cite{jouni,ourdelta,ourpols}, see also Refs.~\cite{withnorbert,Baru_proc}
where the effect of the $\Delta$ in $NN\pi$ was studied within ChPT.

The progress in the theory of isospin conserving pion production in nucleon-nucleon collisions
allowed us to perform a complete leading-order calculation of charge symmetry breaking effects in
$pn \to d\pi^0$ \cite{Ars}, see also Refs.~\cite{KMN,Bolton} for related works.

Given the developments just discussed, one now should be in the position
to investigate two--pion production in $NN$ collisions using the same
formalism. Such calculations do not exist yet, although in the pioneering
works of Refs.~\cite{ruso,ruso2} some constraints from chiral symmetry were already
implemented.  A more recent phenomenological analysis of this reaction can also be found in Ref.~\cite{cao}.
For two-pion production in $NN$ collisions at threshold, the center-of-mass momentum of
the initial nucleons is necessarily larger by a factor of $\sqrt{2}$
compared to the single-pion production.
Thus we now have
\Be
p_{\tiny{\rm thr}}^{(2)}= \sqrt{2m_\pi  M_N}\approx 510 \ {\rm MeV} \ .
\Ee
 With this large momentum scale, the value of the
expansion parameter is \Be \chi=\frac{p_{\tiny{\rm thr}}^{(2)}}{M_N}\approx 0.54 \Ee
and one may question the applicability of ChPT for the two-pion production case.
However, while
the expansion may converge slowly, it is still meaningful at least close to the chiral
limit and therefore should also provide some guideline on the hierarchy of
diagrams.
Thus it is still interesting and important to investigate the structure of diagrams
that contribute at the lowest orders.
Note that, since the $NN$ momentum at the threshold for two-pion production is quite
close to the one for single-pion production, for the purpose of power counting below
we identify the two.

The amount of experimental data for $NN\to NN\pi\pi$ has increased greatly in recent
years~\cite{NNpipidata,Kren,Bashkanov,Johanson}  and even a polarized
measurement
exists~\cite{NNpipipol}. However, up to now there are no data available sufficiently
close to threshold to allow for a straightforward comparison with our
leading order calculation of the reaction  $pn\to d\pi\pi$.
To extend our calculation to higher energies the inclusion of the $\Delta$
as an explicit degree of freedom will be necessary (it enters at NLO).
In addition, at NNLO, besides a large number of loops, also the first
$NN\to NN\pi\pi$ counterterms start to contribute. The physics of those is
dominated by heavier baryon resonances --- most prominently the Roper
resonance, as follows from previous phenomenological
studies of the reactions $NN\to NN\pi\pi$~\cite{ruso,ruso2,cao}.
Two pion production and the role of the Roper resonance, in particular, has been studied
extensively in the single-nucleon sector,
namely in the reactions $\gamma N\to \pi\pi N$  and $\pi N\to \pi\pi N$
 both phenomenologically, see, e.g.,  Refs. \cite{schneider,Kamano1,Kamano2,Gomez}
and within ChPT \cite{BKMSgamma,BKMgamma,BKM_pipiN95,Fettes_pipiN,Mobed}
\footnote{Note that for the reaction $\pi N\to \pi\pi N$ within ChPT we refer
to the calculations at order $O(p^3)$ only, see references cited in
\cite{BKMSgamma,BKMgamma,BKM_pipiN95,Fettes_pipiN,Mobed} for numerous other studies.}.
 In particular, ChPT studies  \cite{BKMSgamma,BKMgamma}
predict that the double-neutral-pion photoproduction is significantly enhanced
near threshold as compared to the production of the charged pions due to the
contribution of chiral (pion) loops at NLO ($O(p^3)$). This prediction
was confirmed experimentally in Ref.~\cite{Kotulla}. Also it was found in
Ref.~\cite{BKMSgamma} that the $\Delta$-isobar contribution, which is potentially
important because the $\Delta-$nucleon mass difference is of comparable magnitude
to $2m_{\pi}$, turns out to be insignificant
due to specific cancellations of the individual contributions.
The Roper resonance contributes to this process
at one order higher, i.e. at N$^2$LO ($O(p^4)$). The effect of the Roper resonance
encoded in low-energy constants  was found to be sizable although significantly smaller
than the contribution of the leading chiral loops \cite{BKMgamma}. Similar conclusions about
the role of the  Roper resonance were drawn based on studies of $\pi N\to \pi\pi N$.
In this case, however, the loops were found to be negligible \cite{Fettes_pipiN}
and the leading contribution was provided by tree level terms.
For two-neutral-pion production in $NN$ collisions we expect the production mechanism to be
in accord with the dimensional analysis, i.e it should be dominated by the tree level graphs at LO.
However, an explicit study of loops at NLO will be necessary to support our
uncertainty estimates.

The paper is structured in the following way:
In Sect. II our theoretical framework is introduced.
Results for the reaction $pn \to d\pi^0\pi^0$ are presented and discussed in
Sect. III.
A summary and some concluding remarks are provided in Sect. IV.
Technical details of our calculations are summarized in two Appendices.

\section{Theoretical Formalism}
Our calculations are based on the effective chiral Lagrangian~\cite{Chiral_L_NN1,Chiral_L_NN2,bernard_rev}:
\Be
\CL^{(0)}&=&\frac{1}{2f_\pi^2}[(\boldm{\pi}\cdot\partial_\mu\boldm{\pi})^2
-\frac{1}{4}\boldm{\pi}^2(\partial^\mu\boldm{\pi})^2]+N^\dag\frac{1}{4f_\pi^2}\boldm{\tau}\cdot(\dot{\boldm{\pi}}\times
\boldm{\pi})N \nonumber \\&&
+\frac{g_A}{2f_\pi}N^\dag\boldm{\tau}\cdot\vec{\sigma}\cdot \left
(\vec{\nabla}\boldm{\pi}+\frac{1}{2f_\pi^2}\boldm{\pi}(\boldm{\pi}\cdot
\vec{\nabla}\boldm{\pi})\right ) N\nonumber  +\cdots\\
\CL^{(1)}&=& \frac{1}{8M_Nf_\pi^2}\left (
iN^\dag\boldm{\tau}\cdot(\boldm{\pi}\times
\vec{\nabla}\boldm{\pi})\cdot\vec{\nabla}N+h.c.\right
)-\frac{g_A}{4M_Nf_\pi}[iN^\dag\boldm{\tau}\cdot
\dot{\boldm{\pi}}\vec{\sigma}\cdot \vec{\nabla}N+h.c.] +\nonumber
\\&& \hspace*{-1.5cm}\frac{1}{f_\pi^2}N^\dag\left [ \left
(c_2+c_3-\frac{g_A^2}{8M_N}\right
)\dot{\boldm{\pi}}^2-c_3(\vec{\nabla}\boldm{\pi})^2-2c_1m_\pi^2\boldm{\pi}^2
-\frac{1}{2}(c_4+\frac{1}{4M_N})\epsilon_{ijk}\epsilon_{abc}\sigma_k\tau_c\partial_i\pi_a\partial_j\pi_b\right
] N \nonumber \\ &&+\cdots \Ee
where the superscripts 0 and 1 denote the leading and next-to-leading order Lagrangian, respectively,
$f_\pi$ denotes the pion decay constant in the chiral limit,
$g_A$ is the axial-vector coupling of the nucleon,
$N$ and $\pi$ are the nucleon and pion field, respectively, and
$\vec{\sigma}$ ($\boldm \tau$) denotes the Pauli matrix in spin (isospin) space.
The ellipses in the above equations represent terms
which are not relevant for the present study.

\begin{figure}[htbp] \vspace{0.cm}
\begin{center}
\includegraphics[scale=1.5]{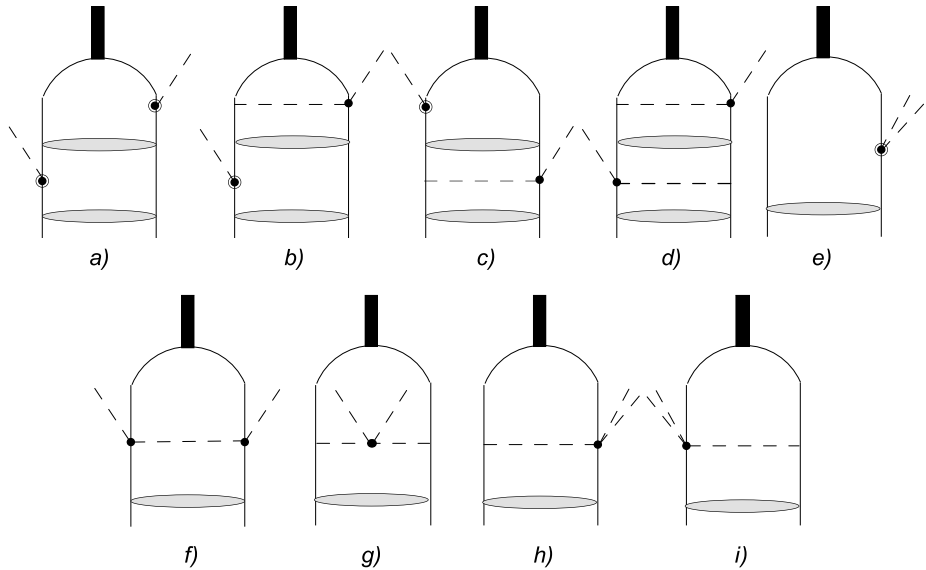}

 \caption{Leading order diagrams for the reaction $pn \rightarrow d \pi\pi$ at
   threshold.
Solid (dashed) lines denote nucleons (pions), filled
ellipses correspond to continuum $NN$ wave functions (including plane wave)
in the initial and intermediate state,  the outgoing black line
denotes the deuteron. Subleading vertices are marked as $\odot$.
 }
 \label{feyn_fig1}
\end{center}
\end{figure}

As mentioned in the Introduction, the system is characterized
by three scales, where we use
 $$m_\pi \ll p \sim p_{\rm thr}^{(1)}\sim p_{\rm thr}^{(2)}\sim \Delta M \ll
\Lambda_\chi \ .$$ The counting rules here are similar to those for one-pion production
used in ~\cite{ch3body,withnorbert,p-wave,lensky1} (for a review see also~\cite{report}). For
the Weinberg-Tomozawa (WT)  $\pi N\to \pi N$ vertex, as discussed in detail in
Ref.~\cite{lensky1}, the leading term should be proportional to $2\omega_\pi$
where $\omega_\pi$ is the energy of the outgoing (on-shell) pion. The diagrams
that contribute at the lowest order are shown in Fig.~\ref{feyn_fig1} and
start to contribute at order $\chi \sim m_\pi/M_N$.
Obviously, the number of diagrams is significantly larger
for two-pion production than for one-pion production already at LO --- after all the
latter appears as building block of the former in diagrams a)--d).
For Fig.~\ref{feyn_fig1}d), we note that the naive dimensional analysis
 shows that this diagram contributes at next--to--leading order. However, due
to the existence of the two--nucleon cut in the intermediate state, this
diagram is promoted to leading order following the very same logic adopted by
Weinberg~\cite{Weinberg:1991um}. In  addition, Fig.~\ref{feyn_fig1} e)--i) show
genuine two--pion production diagrams.
Note that the only non--resonant diagrams included in the earlier studies
of Refs.~\cite{ruso,ruso2} are those depicted in the second line of
Fig.~\ref{feyn_fig1}, whereas in Ref.~\cite{cao} such diagrams are not considered at all.

Another potentially important class of contributions arises from diagrams
where one of the nucleons is excited to the $\Delta(1232)$-resonance.
However, in the reaction considered here the $N\Delta$ system can only couple
to the $NN$ system in the intermediate state. The $NN$ systems in the
initial and final states have to have isospin zero, cf. below, while the
$N\Delta$ system has always isospin one.
The $\Delta(1232)$ can only
appear when the first pion is emitted and, thus, the power counting
for the $\Delta(1232)$ in $pn\to d\pi\pi$ is similar to the one in single-pion production.
In particular, the $\Delta(1232)$ starts to contribute at NLO due to the fact that
the $\Delta$ propagator is suppressed by $1/p$ as compared to 1/$m_\pi$ in the
nucleon case. Moreover, the aforementioned isospin arguments prevent the potentially
dangerous situation when there is an extremely enhanced $\Delta$-propagator
$(\Delta M-2m_{\pi}-{\rm recoil})^{-1}$ in the $N\Delta$ intermediate state that would
occur from a $\Delta$ excitation prior to the first pion emission.
It is interesting to note that a similar observation  was made in Ref.~\cite{BKMSgamma}
in the reaction $\gamma N\to \pi\pi N$ where the corresponding term
$(\Delta M-2m_{\pi})^{-1}$ did not show up in the denominator because of an exact
cancellation with the same term in the numerator.

Once the production operators are constructed, they
should be convoluted with $NN$ wave functions to account for the nonperturbative
character of $NN$ interaction~\cite{Weinberg:1991um}. In principle, the $NN$ wave
functions should be calculated on the same basis as the production operators. However, up to
now  ChPT $NN$ interactions are only available for energies below the one-pion production
threshold~\cite{NN1,NN2,NN3}.
Therefore, in this work we adopt the so-called hybrid approach that
consists in the convolution of the production operator calculated within ChPT
with phenomenological $NN$ wave functions from realistic $NN$ models,
in particular from the CCF model~\cite{ccf}.
We utilize also wave functions generated from the CD-Bonn $NN$ potential
\cite{cdbonn} so that we can examine in how far our results depend on
the specific choice of the $NN$ wave functions.
The $\Delta(1232)$ resonance is included as explicit degree of freedom in
the construction of the CCF potential~\cite{ccf} as well as in
an extended version of the CD-Bonn potential~\cite{cdbonn_delta}.
This will allow us to extend in a future work the present calculation to higher orders
in a straightforward manner.
The ellipses in the Feynman diagrams of Fig.~\ref{feyn_fig1} refer to the $NN$ wave functions
and represent either the Initial State Interaction~(ISI) or the Intermediate State
Interaction~(ImSI) of nucleon pairs supplemented with the plane wave terms.

Based on the Lagrangian given above, it is straightforward to
calculate the amplitudes corresponding to the diagrams shown in
Fig.~\ref{feyn_fig1}. Furthermore, since we are only interested in the energy
region very near to threshold, only a few partial waves contribute to the reaction amplitude.
At threshold, the $\pi\pi$ system is in a relative S-wave and
its isospin can be either zero or two as required by the symmetry of a system of identical
bosons. However, since the isospin of the initial $NN$ system can only be
zero or one and the deuteron is an isosinglet the isospin of the produced $\pi\pi$
pair has to be zero and, consequently, also the isospin of the initial $NN$ state is zero.
Then the conservation of parity and angular momentum further requires that the total
angular momentum of the system should be one and the $NN$ pair in the initial
state can only be in the $^3S_1$ or $^3D_1$ partial waves. The
Pauli principle is also satisfied in this case. The
explicit expressions for the amplitudes of these two partial waves
can be found in Appendix A.

\section{Results and Discussion}

\begin{table}[t!]
\begin{center}
\caption{The values of the  amplitudes for the individual  diagrams (in units of  $10^{-2}{\times}$MeV$^{-1}$). }
\includegraphics{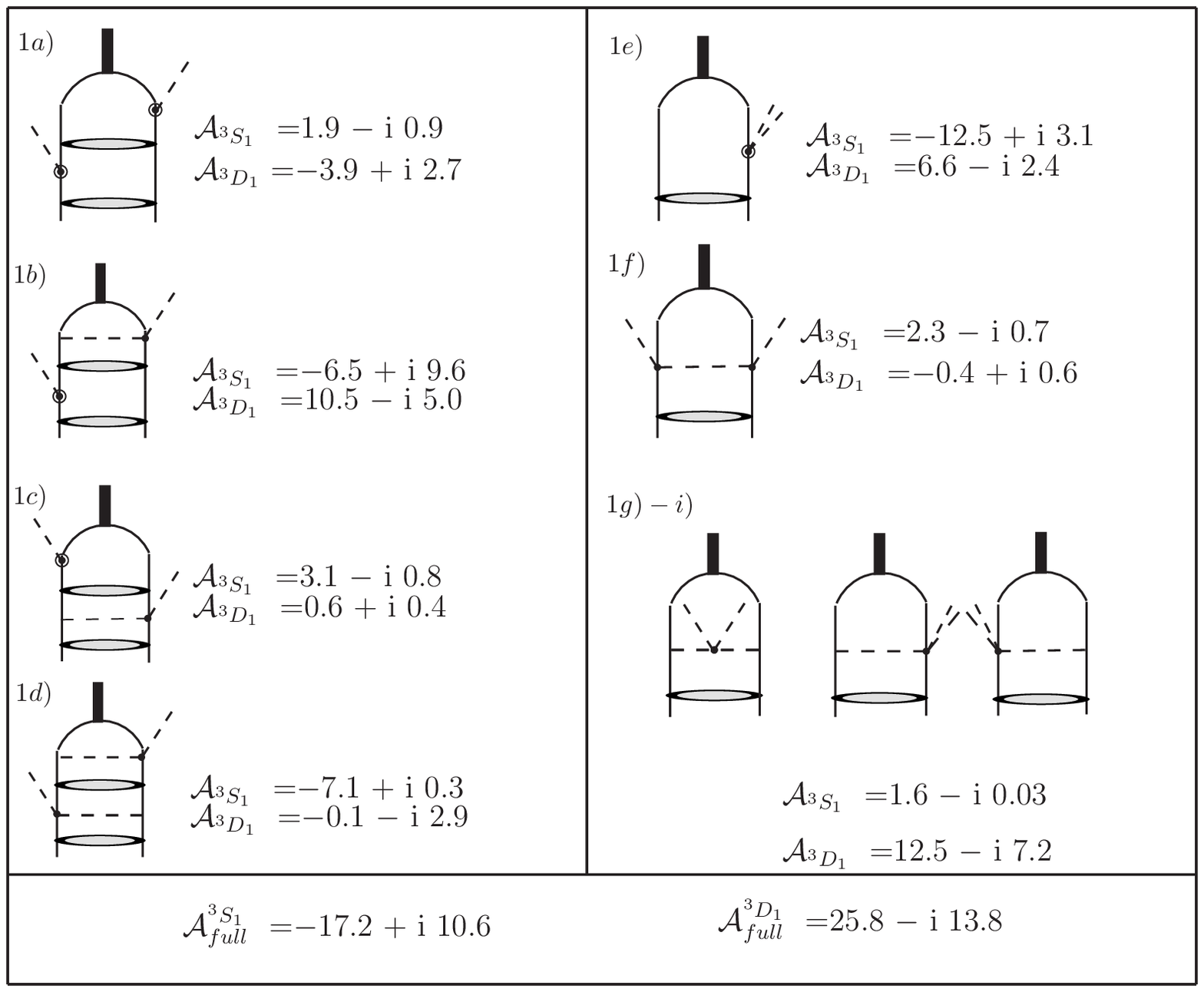}
\label{fullamp}
\end{center}
\end{table}

\begin{table}[t!]
\begin{scriptsize}
\begin{center}
\caption{Values of the individual amplitudes for the $^3S_1$ partial wave (in   $10^{-2}{\times}$MeV$^{-1}$).}
\label{3s1}
\begin{tabular}{|p{4.cm}|c|c|c|c|c|c|c|}
\hline &   \multicolumn{1}{c|}{ \makecell{ Born\\ $A_0$} }& \multicolumn{2}{c|}{\makecell{Initial state \\
Interaction\\$A_{ISI}$ }} & \multicolumn{1}{c|}{ \makecell{Intermediate \\ state \\
Interaction\\$A_{ImSI}$} } & \multicolumn{2}{c|}{\makecell{Initial State Int.
+ \\ Intermediate state Int.\\$A_{ISI+ImSI}$}} & sum \\
\cline{3-4}\cline{6-7}
  & &$^3S_1\rightarrow ^3S_1$ &$^3S_1\rightarrow ^3D_1$  & &$^3S_1\rightarrow ^3S_1$ & $^3S_1\rightarrow ^3D_1$ &\\ \hline
 1a)\parbox[c]{2.3cm}{\epsfig{file=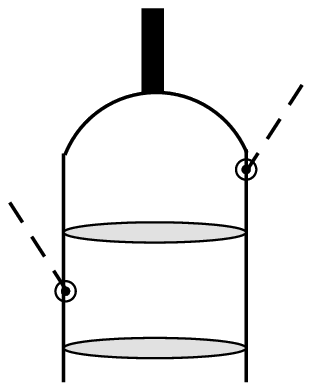, height=1.2cm}}      &$11.6$ &
 $-3.4-i2.8$
 &$-1.6-i0.2$&$-7.8+i0.1$&$2+i1.9$&$1.1+i0.1$&$1.9-i0.9$\\
 \hline
 1b)\parbox[c]{2.3cm}{\epsfig{file=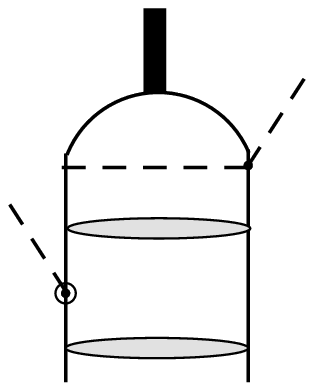, height=1.2cm}}      &$-23.9$ &
 $4+i7.3$
 &$3+i$&$14+i7.5$&$-1.8-i5$&$-1.8-i1.2$&$-6.5+i9.6$\\
 \hline
1c)\parbox[c]{2.3cm}{\epsfig{file=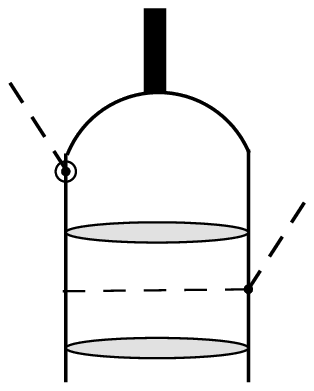, height=1.2cm}}      &$13.3+i3.4$ &
 $-5.5-i3.5$
 &$1.5+i0.2$&$-8.8-i3.2$&$3.4+i2.7$&$-0.8-i0.4$&$3.1-i0.8$\\
 \hline
1d)\parbox[c]{2.3cm}{\epsfig{file=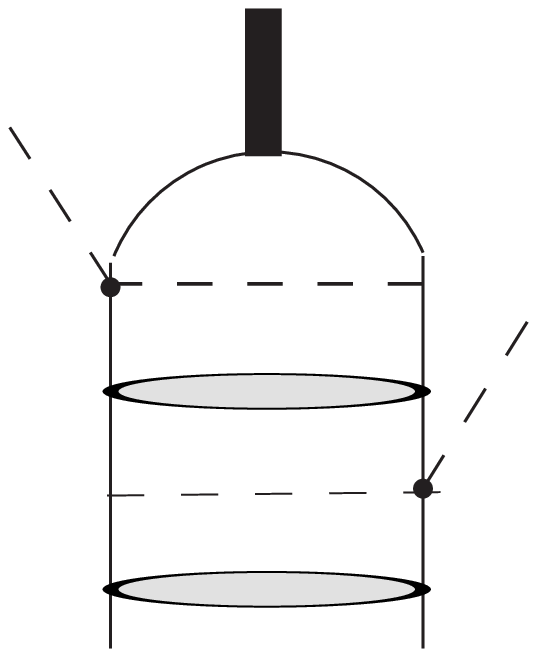, height=1.2cm}}      &$-22.7-i14.5$ &
 $7.5+i10$
 &$-2.1-i2$&$13+i13$&$-3.7-i7.5$&$0.9+i1.3$&$-7.1+i0.3$\\
 \hline
1e)\parbox[c]{2.45cm}{\epsfig{file=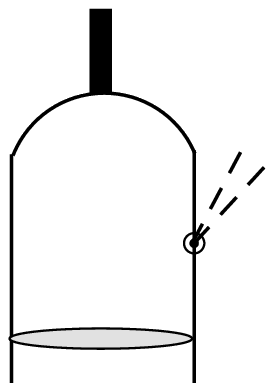, height=1.2cm}}      &$-14.4$ &
 $-6.2+i3.4$
 &$8.1-i0.3$&$-$&$-$&$-$&$-12.5+i3.1$\\
 \hline
1f)\parbox[c]{2.3cm}{\epsfig{file=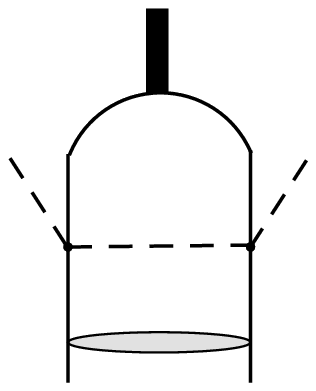, height=1.2cm}}      &$3.5$ &
 $-1.6-i0.6$
 &$0.4-i0.1$&$-$&$-$&$-$&$2.3-i0.7$\\
 \hline
1g)-i) \parbox[c]{2.3cm}{ \epsfig{file=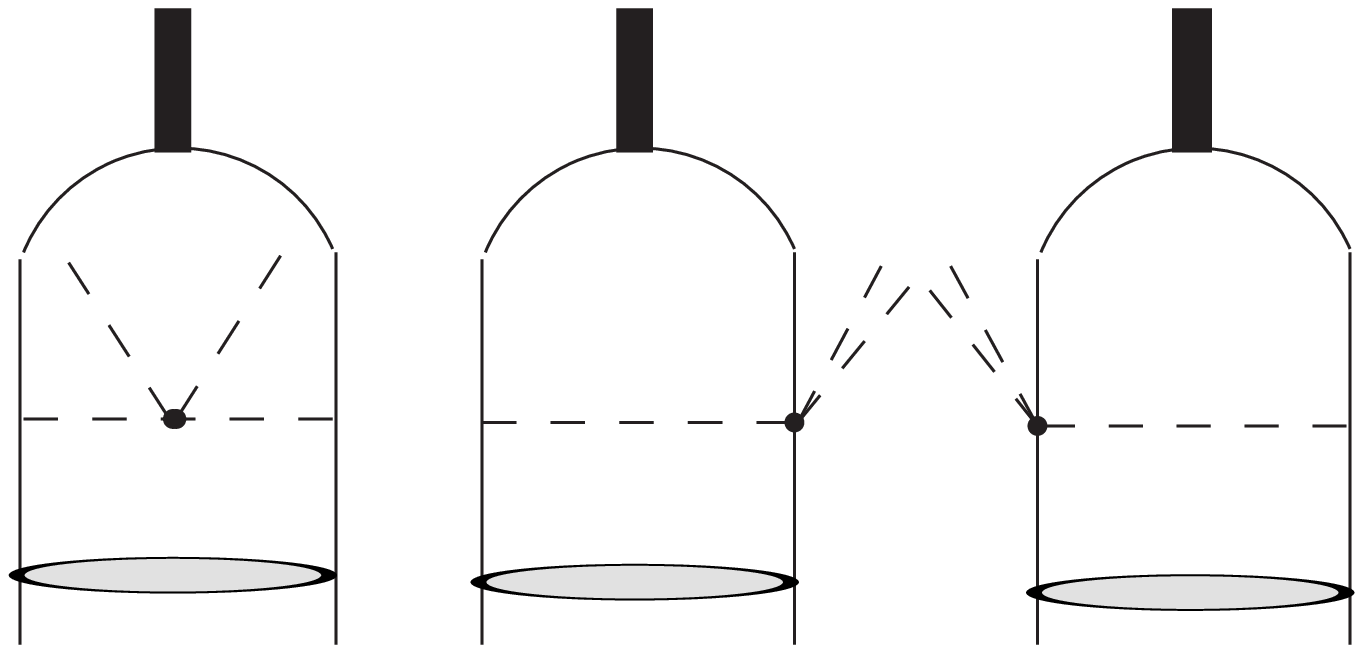, height=1.15cm}}      &$-4.6$ &
 $2+i0.6$
 &$4.2-i0.6$&$-$&$-$&$-$&$1.6-i0.03$\\
 \hline
\end{tabular}
\end{center}
\end{scriptsize}
\end{table}

\begin{table}[t!]
\begin{scriptsize}
\begin{center}
\caption{Values of the individual amplitudes for the  $^3D_1$ partial wave (in  $10^{-2}{\times}$MeV$^{-1}$).}
\label{3d1}
\begin{tabular}{|p{4.cm}|c|c|c|c|c|c|c|}
\hline & \multicolumn{1}{c|}{ \makecell{ Born\\ $A_0$} }& \multicolumn{2}{c|}{\makecell{Initial state \\
Interaction \\$A_{ISI}$}} & \multicolumn{1}{c|}{ \makecell{Intermediate \\ state \\
Interaction\\$A_{ImSI}$} } & \multicolumn{2}{c|}{\makecell{Initial State Int.
+ \\ Intermediate state Int.\\$A_{ISI+ImSI}$}} & sum \\
\cline{3-4}\cline{6-7}
  & &$^3D_1\rightarrow ^3D_1$ &$^3D_1\rightarrow ^3S_1$  & &$^3D_1\rightarrow ^3D_1$ & $^3D_1\rightarrow ^3S_1$ &\\ \hline
 1a)\parbox[c]{2.3cm}{\epsfig{file=a.eps, height=1.2cm}}      &$-8.2$ &
 $0.5+i3.3$
 &$-2.4+i0.4$&$5.5-i0.1$&$-0.4-i1.7$&$1.1+i0.8$&$-3.9+i2.7$\\
 \hline
 1b)\parbox[c]{2.3cm}{\epsfig{file=c.eps, height=1.2cm}}      &$16.9$ &
 $1.2-i5.2$
 &$4.1+i4.8$&$-9.9-i5.3$&$-0.9+i3.1$&$-0.9-i2.4$&$10.5-i5$\\
 \hline
1c)\parbox[c]{2.3cm}{\epsfig{file=b.eps, height=1.2cm}}      &$4.4+i2.4$ &
 $-1.5-i1.9$
 &$-3.1+i1.5$&$-1.7-i2.3$&$0.5+i1.1$&$2-i0.4$&$0.6+i0.4$\\
 \hline
 1d)\parbox[c]{2.3cm}{\epsfig{file=wt_box.eps, height=1.2cm}}      &$-3.8-i10$ &
 $-0.04+i4$
 &$6.5+i0.8$&$0.7+i5.4$&$0.1-i2.1$&$-3.6-i$&$-0.1-i2.9$\\
 \hline
 1e)\parbox[c]{2.45cm}{\epsfig{file=e.eps, height=1.2cm}}      &$37.9$ &
 $-5.1-i19.4$
 &$-26.2+i17$&$-$&$-$&$-$&$6.6-i2.4$\\
 \hline
 1f)\parbox[c]{2.3cm}{\epsfig{file=d.eps, height=1.2cm,clip=}}      &$1.6$ &
 $-0.6-i0.6$
 &$-1.4+i1.2$&$-$&$-$&$-$&$-0.4+i0.6$\\
 \hline
  1g)-i) \parbox[c]{2.3cm}{\epsfig{file=f+g+h.eps, height=1.2cm,clip=}}      &$17.9$ &
 $-5.1-i7.1$
 &$-0.3-i0.1$&$-$&$-$&$-$&$12.5-i7.2$\\
 \hline
\end{tabular}
\end{center}
\end{scriptsize}
\end{table}

As discussed in the introduction, some leading order diagrams depicted in
Figs.~\ref{feyn_fig1}a)-~\ref{feyn_fig1}e) are missing in
Ref.~\cite{ruso}.  First, we want to discuss the role
of the diagrams ~\ref{feyn_fig1}a)-~\ref{feyn_fig1}d). As can be seen from the figure,
these diagrams are closely related to the one-pion production
case. There the leading order contributions come from
the direct pion emission and the rescattering term. Thus, one may expect that the
amplitudes  ~\ref{feyn_fig1}a)-~\ref{feyn_fig1}d) reveal a
pattern similar to the one-pion production case.  For example, one may guess that the amplitude
of Fig.~\ref{feyn_fig1}a) should be much smaller than the one of
Fig.~\ref{feyn_fig1}b) due to the fact  that for single-pion production the rescattering term
dominates over the direct one by an order of magnitude.  However, it should be noted that in our case
the $NN$ amplitude in the intermediate state is fully off-shell and it turns out that 
the real part of diagram b) is larger only by a factor of about 3 as compared to the
one of diagram a) --- cf. Table~\ref{fullamp}.
In addition, due to the fact that the $NN$ amplitudes for the ISI
appear in coupled channels ($^3S_1$-$^3D_1$)
the production amplitudes, as given in Table~\ref{fullamp}, do not have the same phase.
 This fact makes the interference between the individual amplitudes very
different from the one-pion production case and a direct comparison becomes
difficult. The role of the $NN$ interaction in the initial and intermediate states can be
understood from looking at the individual contributions to the two-pion production
amplitudes listed in Tables ~\ref{3s1} and ~\ref{3d1}.
We find that the
values of these contributions are basically comparable with each other, as
expected from the power counting. The values of the amplitudes of Fig.~\ref{feyn_fig1}b) and
Fig.~\ref{feyn_fig1}d) are about a factor of two to three larger than the other two
amplitudes \ref{feyn_fig1}a) and \ref{feyn_fig1}c),
which is still a reflection of the dynamics governing the one-pion production.
We observe that the resulting values of the amplitudes ~\ref{feyn_fig1}a)-~\ref{feyn_fig1}d)
are similar in size to those of ~\ref{feyn_fig1}f)-~\ref{feyn_fig1}i)
considered in previous studies \cite{ruso,ruso2} and, thus, are important.
We also find that the $NN$ interaction in the initial and intermediate states plays
an important role in the calculation, especially for the diagrams ~\ref{feyn_fig1}a)-~\ref{feyn_fig1}d).
In particular, one can find from Tables~\ref{3s1} and \ref{3d1}
 that the inclusion of the $NN$ interaction in the ISI or the ImSI
 generally reduces the magnitude of the amplitudes as compared to
the Born amplitude and that  the cancellation among the individual amplitudes of each particular
diagram is significant.  The largest cancellation takes place between the amplitudes
with and without the ImSI, i.e. between ($\CA_0$ and $\CA_{ImSI}$) and ($\CA_{ISI}$ and $\CA_{ISI+ImSI}$)
where the amplitudes are defined in  Appendix A, see also Tables~\ref{3s1} and \ref{3d1} .
For example, the value of the S-wave amplitude of Fig.~\ref{feyn_fig1}a) becomes a factor of two
smaller than the Born amplitude after including the ISI. When the ImSI is also considered, the full
amplitude is further reduced so that it is finally a factor of six smaller than the Born
amplitude. This observation  shows explicitly that the ISI and the ImSI are very
important quantitatively. Since they also strongly influence the phase
of the amplitudes, their proper inclusion is compulsory, especially
for studies of polarization observables.

 The strength of the contact terms, that appear in Fig.~\ref{feyn_fig1}e), is given
by the low energy constants $c_i$ that can be determined from $\pi N$
 scattering.  Because we are only interested in the amplitudes at threshold,
 we only need the values of $c_1$, $c_2$ and $c_3$.  In this work, we take the
 values from Ref.~\cite{epelbaum} where the values of $c_i$'s are obtained by
 fitting $\pi N$ threshold parameters. To the order we are working, three
 solutions are offered in that paper. One corresponds to the results without
 considering the $\Delta(1232)$ explicitly, and the other two are results where
 the $\Delta(1232)$ is included but with different choices of the $\pi N\Delta$
 coupling constant $h_A$. It is well known that $c_2$ and $c_3$ can be largely
 accounted for by the contribution from the $\Delta(1232)$~\cite{bernard}, and
therefore the
 values of $c_2$ and $c_3$ can change significantly between extractions with
 and without inclusion of the $\Delta(1232)$. One may expect that using
 different sets of parameters will affect the value of the amplitudes of
 Fig.~\ref{feyn_fig1}e) strongly. However, it is interesting to note that the
 amplitude of Fig.~\ref{feyn_fig1}e) is independent of those choices, because
 it only depends on the value of the linear combination
\begin{equation}
\label{cicomb}
 2m_\pi^2 c_1+q_1^0q_2^0(c_2+c_3-g_A^2/(8M_N)) \ ,
\end{equation}
with $q_1^0=q_2^0=m_\pi$ for the pion energies
in the kinematics relevant here.
From Table~1 of Ref.~\cite{epelbaum} one can see that the values
of $c_1$ and $c_2+c_3$ are not affected by the $\Delta(1232)$ up to order
$O(p^2)$. Thus, although the values of $c_2$ and $c_3$ can significantly
change individually by considering the $\Delta(1232)$ explicitly, the sum
of them is not affected up to order $O(p^2)$, as shown explicitly in
Ref.~\cite{bernard}. Hence at LO the contribution from
contact terms are not influenced by the $\Delta(1232)$, which is also
consistent with the power counting used. It is also interesting to note that
the linear combination of low energy parameters displayed in
Eq.~(\ref{cicomb}) is large, for the individual terms interfere constructively
--- this is in contradistinction to $\pi N$ scattering, where $q_1^0=-q_2^0=m_\pi$
holds at threshold and then physics is governed by the very small
isoscalar scattering length~\cite{bernard}, see Ref.~\cite{pid_isospinviol} for
a recent accurate extraction of $a^+$ from pionic atoms.
Thus, once the reaction
$pn\to d\pi^0\pi^0$ can be studied with sufficient accuracy, it might be a
good source of information for the $c_i$ individually.  Note also
that in Ref.~\cite{epelbaum} the 1/$M_N$ corrections are not included explicitly
 in the fits to $\pi N$ data but absorbed effectively into the LECs $c_i$. To be consistent
with their treatment we also drop  the  1/$M_N$ term in Eq.~(\ref{cicomb}).
Furthermore, the residual combination $c_2+c_3$
which enters Eq.~(\ref{cicomb}) is totally determined by the $S$-wave
threshold parameters.
Numerically the result for diagram~\ref{feyn_fig1}e)
turns out to be one of the largest individual contributions --- cf. Table~\ref{fullamp}.

The diagrams\footnote{We checked,
using the methods of Ref.~\cite{withandreas}, that the sum of the
contributions from Figs.~\ref{feyn_fig1}g)-~\ref{feyn_fig1}i) is independent
of the choice of the pion field, as required by field-theoretic
consistency.}
of Figs.~\ref{feyn_fig1}f)-~\ref{feyn_fig1}i)
are included in the previous study of Ref.~\cite{ruso}, except for
the initial state interaction which is not considered in that work. However, as can
be seen from Tables~\ref{3s1} and \ref{3d1}, the consistent inclusion of the ISI changes
the value of the amplitudes \ref{feyn_fig1}f)-~\ref{feyn_fig1}i) significantly and is thus important.

Besides the interference between the individual amplitudes of each particular diagram, the
interference between the contributions from different diagrams is also very significant
as can be seen from Tables~\ref{3s1} and \ref{3d1}.
In order to examine the sensitivity of our results and
especially of the interference pattern to the employed $NN$ interaction
we also performed calculations with a different $NN$ model, namely with
the CD-Bonn model~\cite{cdbonn}.
We found that the values of the individual amplitudes vary in a
reasonable range (up to 30\%) due to differences in the $NN$ wave functions
whereas the sum of all amplitudes varies just  by a small amount (around 10\%).
This variation should be partly absorbed by the contribution of the
$(N^\dagger N)^2\pi\pi$ contact term at N$^2$LO.

\begin{figure}[t!] \vspace{0.cm}
\begin{center}
\includegraphics[scale=0.5]{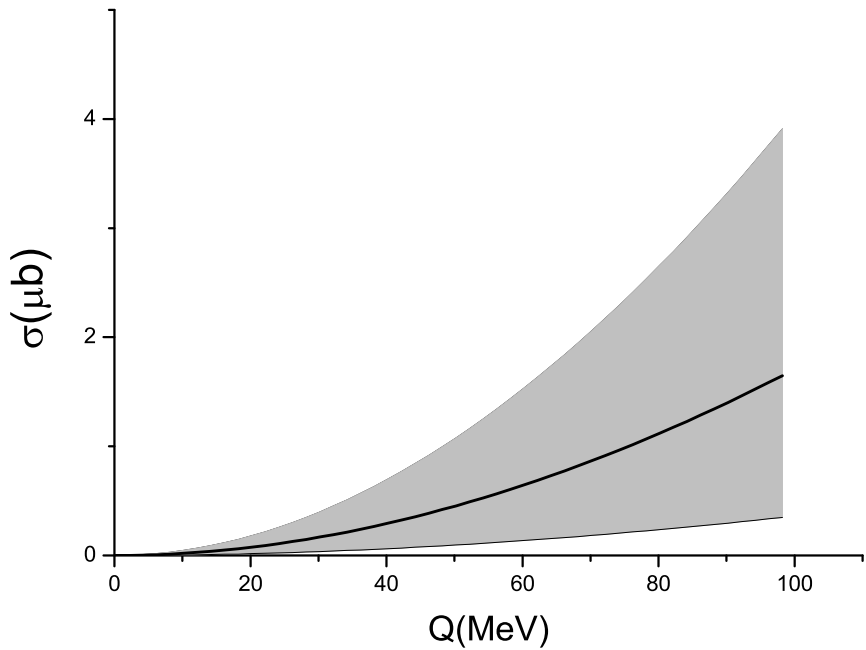}
 \caption{Result of our ChPT calculation for the cross section of $pn\to d\pi^0\pi^0$
 based on the assumption of a constant production
amplitude. The shaded area corresponds to the uncertainty estimate
based on a dimensional analysis.
}
 \label{xsec}
\end{center}
\end{figure}

 To check the convergence of chiral expansions based on counting rules, it is
 necessary to calculate explicitly subleading contributions. As those
 calculations are not available yet, we may try to compare the results from
 the leading order contributions to experimental observables: our result
 should be within 50\% of the data, given the relatively large value of the
 expansion parameter.  The applicability of our $s$-wave calculation is
 restricted to the near-threshold regime and, therefore, data close to the
 threshold are needed for a sensible comparison of theory with
 experiment. Unfortunately, at present the empirical information for the
 $d\pi\pi$ channel is very scarce. Only data down to excess energies $Q=70$ MeV
($Q=\sqrt{s}-\sqrt{s_{\rm thr}}$) are available and the uncertainties are
 still very large. Thus, it is difficult to draw any conclusions about the
 convergence in the present work. In order to provide some comparison of our
 calculation with the existing data, we assume that the matrix element is
 constant and therefore take the energy dependence as originating only from
 the three--body phase space. Our result is roughly a factor of two smaller
 than the central value of the data point at $Q=70$ MeV. We want to stress,
 however, that the uncertainties of the experiment in question is quite large
 (even a factor of 25 larger than the data themselves). For future reference,
 in Fig.~\ref{xsec} we show the central value for the cross section predicted
 under the assumption that the matrix element is a constant and that no higher
 partial waves contribute. We also indicate the uncertainty band based on a
 dimensional analysis.

\section{Summary}

We presented a complete leading order calculation for the
reaction $pn\to d\pi^0\pi^0$ at threshold within chiral perturbation theory.
There is no free parameter in the calculation at this order.
We included various additional diagrams as compared to previous
investigations~\cite{ruso,ruso2}
and we also made several technical improvements.
Our most important findings, summarized in Table~\ref{fullamp}, are
\begin{itemize}
\item all diagrams evaluated are of similar importance;
\item there are sizable interferences between the individual contributions;
\item the accurate inclusion of the $NN$ interaction in both
intermediate as well as initial states is very important.
\end{itemize}
We also stress that for two--pion production the expansion
parameter
$$
 \chi=\frac{p_{\tiny{\rm thr}}^{(2)}}{M_N}\approx 0.54 \ ,
$$ where $p_{\tiny{\rm thr}}^{(2)}$ denotes the initial $NN$ momentum at the
two-pion production threshold, is rather large.
It is therefore important to calculate higher order contributions to check the rate of
convergence. We argue in this paper that
despite of the proximity of the \Del to the $\pi\pi N$ threshold the
potentially most important $N\Delta$ intermediate state is not allowed for
the reaction $pn\to d\pi\pi$ because of isospin conservation.
Intermediate states containing the \Del must be of the kind $\pi N\Delta$,
i.e. can only occur after one-pion emission.
Therefore, the role of the \Del resonance in the reaction considered
is expected to be analogous to that in one-pion production. In particular, the \Del
starts to contribute at next-to-leading order.
Furthermore, it is  known from phenomenological studies of $NN\to NN\pi\pi$~\cite{ruso,ruso2}
that the Roper resonance can play a significant role already near threshold and that it
will become even more important when considering a larger range of energies.
However, this
resonance is not included explicitly in the present study.  One may expect
that its contribution is absorbed into some low energy constants. The
contribution of the Roper resonance to the $c_i$ parameters, that scale the
strength of the leading isoscalar $\pi$N scattering, has been discussed, e.g.,
in Ref.~\cite{bernard}. It seems that it plays only a minor role here. Similar
conclusions were drawn from systematic studies within ChPT of the double-pion
photoproduction process \cite{BKMSgamma,BKMgamma} and the reaction $\pi N\to
\pi\pi N$ \cite{BKM_pipiN95,Fettes_pipiN,Mobed} near threshold.  In
particular, for the reaction $\gamma p\to \pi^0\pi^0 p$ the contribution of
the Roper was found to be rather moderate as compared to the large
contribution of chiral loops \cite{BKMgamma}.  A recent model calculation of
$\pi N\to \pi\pi N$ by S.~Schneider et al.~\cite{schneider} suggests also that
the Roper resonance plays a rather minor role.  On the other hand, the Roper
might contribute significantly to the $(N^\dagger N)^2\pi\pi$ counterterms, which
enter at NNLO in a ChPT calculation of $NN\to NN\pi\pi$.  Based on the
discussions above, one can not expect that our current calculation can
describe the experimental data well at higher energies. However, the presented
calculation provides an estimate for the contribution of the non--resonant
background near threshold.  It therefore forms a basis for future studies and
is thus a precondition to extract reliable information on the Roper resonance
from near-threshold experiments.

\acknowledgments We thank  U.-G.~Mei{\ss}ner and H. Clement for useful comments.  B.C.Liu
acknowledges the supports from the Helmholtz-China Scholarship Council Exchange
Program and the National Natural Science Foundation of China under Grants
No. 10905046. This work was supported in parts by funds from the Helmholtz Association
(grants VH-NG-222, VH-VI-231), by the DFG (grants SFB/TR 16 and
436 RUS 113/991/0-1), by the EU HadronPhysics2 project, and by the RFFI. The
work of V.B. was supported by the State Corporation of the Russian Federation, ``Rosatom''.

\appendix
\section{ Reaction amplitudes}
In this appendix, we present expressions for the amplitudes that we consider in this work.
To obtain the partial wave amplitudes, we follow the technics developed in Ref.~\cite{lensky_thesis}.

 For the reaction $NN\rightarrow d\pi\pi$ at threshold, the initial state can
 only be in the $^3S_1$ or $^3D_1$ partial waves and the total isospin is zero. Thus,
 the amplitudes for $NN \rightarrow d \pi\pi$ at threshold can be
 written in the general form  \Be \CM = \left ( \vec{\epsilon}^*\cdot
 \vec{\CS}^{[^3\!S_1]}\CA_{full}^{^3\!S_1} + \vec{\epsilon}^*\cdot
 \vec{\CS}^{[^3\!D_1]}\CA_{full}^{^3\!D_1} \right ) \cdot  (\vec{\pi}\cdot\vec{\pi})\CI_0\ , \Ee where
 $\CS^{[^3\!S_1]}_i=\chi_2^T\sigma_2\frac{\sigma_i}{\sqrt{2}}\chi_1$ and
 $\CS^{[^3\!D_1]}_i=\frac{3}{2}(\delta_{ij}-\frac{1}{3}\hat{n}_i\hat{n}_j)\cdot\chi_2^T\sigma_2\sigma_j
 \chi_1$ denote the normalized spin-orbit structure of the initial nucleons for
the  $^3\!S_1$ and $^3\!D_1$ partial waves, respectively,
 $\CI_0=\varphi_2^T\frac{\tau_2}{\sqrt{2}}\varphi_1$ denotes the isospin
 structure of the initial states, $\CA_{full}^{^3\!S_1}$ and
 $\CA_{full}^{^3\!D_1}$ are the corresponding partial wave amplitudes, $\vec{\pi}$ is the isospin wave function
 of outgoing pions and
 $\vec \varepsilon$ is the deuteron polarization vector. Here
 $\chi_1$($\chi_2$) and $\varphi_1$($\varphi_2$) are spinors of the initial
 nucleons in spin and isospin space, respectively, and $\hat{n}$ denotes the unit
 vector of the relative momentum of the initial nucleons. The expressions for
 $\CA_{full}^{^3\!S_1(^3\!D_1)}$ are obtained by projecting the amplitudes
 that contribute to a given order into corresponding partial waves using the
 technics developed in Ref.~\cite{lensky_thesis}. To leading order, the
 relevant diagrams in our work are shown in Fig.~\ref{feyn_fig1}. Each
 diagram in Fig.~\ref{feyn_fig1}  represents a set of diagrams, in which
 possible $NN$ interaction in the initial and intermediate $NN$ states are
 considered. For example, the partial wave amplitudes of
 Figs.~\ref{feyn_fig1}a)-d) originate from several parts as
 follows: \be
 \CA^{PW}=\CA_0^{PW}+\CA^{PW}_{ISI}+\CA^{PW}_{ImSI}+\CA^{PW}_{ISI+ImSI} \ , \ee
 where the index '0' stands for the Born diagram, and 'ISI', 'ImSI' or 'ISI+ImSI'
 correspond to  the  initial state interaction, intermediate state interaction, or both
 initial  and intermediate state interactions  included in
 the amplitudes, respectively. 'PW' represents the two possible partial waves
 in the initial state, which can be either $^3S_1$ or $^3D_1$. For
 Figs.~\ref{feyn_fig1}e)-i) the partial wave amplitudes  contain only two parts,
 namely $\CA_0^{PW}$ and $\CA^{PW}_{ISI}$ because there is no intermediate
 state interaction.

To compute diagrams with the $NN$ interaction in the initial and intermediate
states we take the $NN$ scattering amplitudes from some potential models.
In our calculation, instead of the commonly used $\CT $
matrix, the $\CM$ matrix is used. These quantities are related by $\CM=-8\pi^2
M_N^2 \CT$. For the initial state
interaction $\CM^{PW\rightarrow PW'}_{NN}$  denotes the $NN$
half-off-shell $\CM$ matrix, where $PW\rightarrow PW'$ represents the
transition from the partial wave ``$PW$'' to ``$PW'$ ''  with
``$PW$'' and ``$PW'$ '' being $^3\!S_1$ or $^3\!D_1$ in our case. For the
intermediate state interaction, the only possible transition is $^3\!P_1
\rightarrow ^3\!P_1$, so we just use $\CM_{NN}^{^3\!P_1 \rightarrow ^3\!P_1}$
to denote the fully off-shell $NN$ $\CM$ matrix. For the deuteron-NN vertex we adopt
the notation and structures used in Ref.~\cite{dNN}. We use $u(p)$ and $w(p)$
to denote the $NN$ components of the deuteron wave function corresponding to
S-wave and D-wave respectively. Clearly, the bound (deuteron) wave functions
$u(p)$ and $w(p)$ should be calculated using  the same potential model as the
 $NN$ scattering amplitudes in the continuum state. In this work we use the CCF
model~\cite{ccf} to generate the $NN$ amplitudes and the deuteron wave
functions. Since the $\Delta\Delta$ channel is considered in the CCF model, that can also
couple to the deuteron, the $NN$ part of the
deuteron wave functions, $u(p)$ and $w(p)$, are normalized
as: \be \int \frac{d^3p}{(2\pi)^3} \left ( u(p)^2 + w(p)^2 \right ) = 0.9864 \ .
\ee

In this work, we adopt the following values of the parameters:
$f_\pi=92.4$ MeV, $g_A=1.32$, $m_\pi=139.58$ MeV and $M_N=938.27$ MeV. The loop
integration is regularized using the cut-off method with $p_{\rm cut}=\Lambda_{\rm ChPT}\approx 1$ GeV.
We checked that the dependence of the results on this parameter
is small -- the results change by 10\% when $\Lambda$ is increased to 10 GeV.

\subsection{Amplitudes of Fig.~\ref{feyn_fig1} a)-d)}
\begin{figure}[htbp] \vspace{0.cm}
\begin{center}
\includegraphics[scale=0.7]{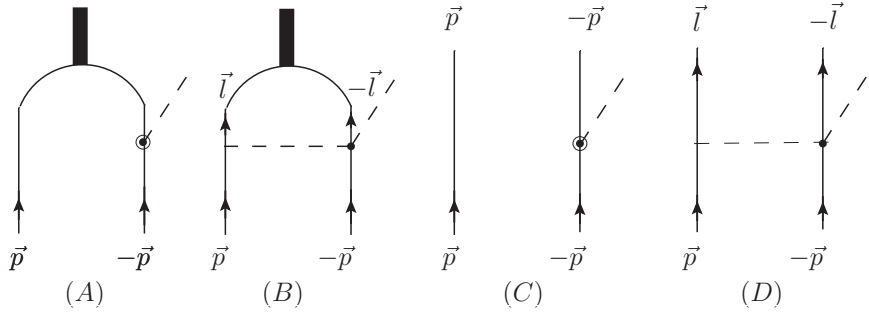}

 \caption{Building blocks for construction the diagrams a)-d) of Fig.~\ref{feyn_fig1}}
 \label{feyn_fig2}
\end{center}
\end{figure}
To get the partial wave amplitudes of
Fig.~\ref{feyn_fig1} a)-d) it is convenient to separate these
diagrams into some building blocks at the position where
the intermediate state interaction may occur, as shown in Fig.~\ref{feyn_fig2}.
Here the diagrams (A) and (B) correspond to the transition from the $^3P_1$ partial
wave that is realized in the intermediate state in the diagrams of Fig.~\ref{feyn_fig1} a)-d)
to the $^3S_1$ deuteron state. Correspondingly, the diagrams (C) and (D) represent the transition
for the $^3S_1-^3D_1$ initial state of the two-nucleon pair to the intermediate $^3P_1$  state.
The full amplitudes can
then be obtained from the expressions for these building blocks by
supplementing propagators, possible $NN$ amplitudes and loop
integrations. The amplitudes of Fig.~\ref{feyn_fig2}(A) and
(B) can be expressed in general as:
\Be \CM&=&\left (\vec{\epsilon}^* \cdot
\vec{\CS}^{[^3\!P_1]} ~ \vec{\CI_1}\cdot \vec{\pi} \right ) \CB \ , \Ee
where
$\pi_i$ is the isospin wave function of the $\pi$ meson in the final state.  Here we introduced
$\CS^{[^3\!P_1]}_i=\frac{\sqrt{3}}{2}\epsilon_{ijk}\hat{p}_k\chi^T_2\sigma_2
\sigma_j \chi_1$ and $\CI_{1,i}=\varphi^T_2 \tau_2\frac{\tau_i}{\sqrt{2}}
\varphi_1$ to denote the normalized spin and isospin structures of the
intermediate $NN$ state, and $\hat{p}=\vec{p}/p$ is the unit vector of
the relative $NN$ momentum in the  intermediate state.
The expressions of $\CB$ for diagrams (A) and (B) are : \Be \CB_A(p)&=&
-\sqrt{\frac{8}{3}}\frac{g_Am_\pi\sqrt{M_N}}{f_\pi}p\left
(u(p)+\frac{w(p)}{2}\right ) \ , \\ \CB_B(p)&=& \frac{4}{\sqrt{3}}\frac{g_Am_\pi
  M_N^{\frac{3}{2}}}{f_\pi^3}\int \frac{l^2 dl}{2\pi^2}\int
\frac{d\Omega_{\hat{p}}}{4\pi}\frac{d\Omega_{\hat{l}}}{4\pi}\frac{1}
     {(\vec{p}-\vec{l})^2+m_\pi^2}\left (
     \frac{p}{4}(w(p)-\frac{u(p)}{\sqrt{2}})\right . \nonumber \\ && \left
     .-\frac{3p}{4}w(p)(\hat p \cdot \hat l
     )^2+\frac{l}{\sqrt{2}}(u(p)+\frac{w(p)}{\sqrt{2}})\hat p \cdot \hat l
     \right ) \ . \Ee
Correspondingly, the expressions of the building blocks for Fig.~\ref{feyn_fig2}(C)
and (D) can be given in the following way:
\Be \CM&=& \left [ (\vec{\CS}^{[^3\!S_1]}\cdot
       \vec{\CS}^{[^3\!P_1]\dag})~~\CB^{^3\!S_1} + (\vec{\CS}^{[^3\!D_1]}\cdot
       \vec{\CS}^{[^3\!P_1]\dag}) ~~\CB^{^3\!D_1} \right ] ~\cdot ~
     (\vec{\CI}_1^\dag \cdot \vec{\pi})\CI_0 \ , \Ee
where $\CB^{^3\!S_1}$ and $\CB^{^3\!D_1}$ for the diagrams (C) and (D) read:
\Be \CB^{^3\!S_1}_C(p)&=&
     \frac{p}{\sqrt{6}}\frac{g_A m_\pi}{M_Nf_\pi} \ , \\ \CB^{^3\!D_1}_C(p)&=&
     -\frac{p}{2\sqrt{3}}\frac{g_A m_\pi}{M_Nf_\pi} \ , \\ \CB^{^3\!S_1}_D(p,l)&=&
     -\frac{1}{\sqrt{6}}\frac{g_Am_\pi}{f_\pi^3}\int
     \frac{d\Omega_{\hat{p}}}{4\pi}\frac{d\Omega_{\hat{l}}}{4\pi}\frac{p\hat p
       \cdot \hat l-l}{(\vec{p}-\vec{l})^2+m_\pi^2} \ , \\ \CB^{^3\!D_1}_D(p,l)&=&
     -\frac{\sqrt{3}}{4}\frac{g_A m_\pi}{f_\pi^3}\int
     \frac{d\Omega_{\hat{p}}}{4\pi}\frac{d\Omega_{\hat{l}}}{4\pi}\frac{l(\hat
       p \cdot \hat l)^2-\frac{2p}{3}\hat p \cdot \hat
       l-\frac{l}{3}}{(\vec{p}-\vec{l})^2+m_\pi^2} \ . \Ee

Given the building blocks shown above and the corresponding $NN$ scattering
amplitudes, it is straightforward to
construct the amplitudes of Fig.~\ref{feyn_fig1} a)-d). For example, the amplitudes
corresponding to diagram \ref{feyn_fig1}a) are:
\Be \CA_0^{a,PW}&=&N_S\cdot \CB_
C^{PW}(p_i)\cdot\frac{M_N}{p_i^2-p_1^2}
\cdot
\CB_A(p_i) \ , \\
\CA_{ISI}^{a,PW}&=&N_S \cdot  \sum_{PW'}^{^3S_1,^3D_1}\int
\frac{p^2dp}{2\pi^2}~\frac{\CM_{NN}^{PW\rightarrow
PW'}(p_i,p)}{(2M_N)^2}\cdot \frac{M_N}{p^2-p_2^2-i\epsilon} \cdot \CB_
C^{PW'}(p)  \nonumber \\
&& \cdot\frac{M_N}{p^2-p_1^2-i\varepsilon}\cdot
\CB_A(p) \ ,
\\
\CA_{ImSI}^{a,PW}&=&N_S \cdot  \CB^{PW}_ C(p_i)
\cdot\frac{M_N}{p_i^2-p_1^2}\cdot \int
\frac{l^2dl}{2\pi^2}~\frac{\CM^{^3P_1\rightarrow
^3P_1}_{NN}(p_i,l)}{(2M_N)^2}  \cdot
\frac{M_N}{l^2-p_1^2-i\epsilon}\cdot
\CB_A(l) \ , \\
\CA_{ISI+ImSI}^{a,PW}&=&N_S\cdot \sum_{PW'}^{^3S_1,^3D_1}\int
\frac{p^2dp}{2\pi^2}~\frac{\CM_{NN}^{PW\rightarrow
PW'}(p_i,p)}{(2M_N)^2}\cdot \frac{M_N}{p^2-p_2^2-i\epsilon}\cdot \CB_
C^{PW'}(p)
\cdot\frac{M_N}{p^2-p_1^2-i\varepsilon}\nonumber
 \\ &&\cdot \int \frac{l^2dl}{2\pi^2}~\frac{\CM_{NN}^{^3P_1\rightarrow ^3P_1}(p,l)}{(2M_N)^2}
\cdot\frac{M_N}{l^2-p_1^2-i\epsilon}\cdot
\CB_A(l), \Ee
where $p_2=p_{\rm thr}^{(2)}$ and $p_1=p_{\rm thr}^{(1)}$.
Here $N_S$ is the symmetry factor which is obtained by considering the
interchange of identical particles in the initial, intermediate and
final states. This factor is the same as the number of Feynman diagrams
one can get if we consider different ways to contract the operators
in the initial, intermediate and final states. For diagrams \ref{feyn_fig1}a)-d)
$N_s$=$ 8 \cdot 2 \cdot \frac{1}{ \sqrt{2}}$. Here 8 comes from the eight ways to
contract the nucleon field operators in the initial and intermediate states; 2 is to account
the interchange of two identical pions in final states;
$\frac{1}{\sqrt{2}}$ is from the deuteron vertex. It should be noted that in
this way the possibility to produce pions from both nucleon lines is included.
The corresponding expressions
of   Fig.~\ref{feyn_fig1}b) can be obtained by changing $\CC$ to $\CD$ in the above equations.

The  amplitudes of Figs. ~\ref{feyn_fig1}c) and ~\ref{feyn_fig1}d) can be obtained in a similar manner.
 The amplitudes of Fig.~\ref{feyn_fig1}c) are
\Be \CA_0^{c,PW}&=&N_S\cdot \int \frac{l^2dl}{2\pi^2}~ \CB_
D^{PW}(p_i,l)\cdot\frac{M_N}{l^2-p_1^2-i\epsilon}
\cdot
\CB_A(l) \ , \\
\CA_{ISI}^{c,PW}&=&N_S\cdot  \sum_{PW'}^{^3S_1,^3D_1}\int
\frac{p^2dp}{2\pi^2}~\frac{\CM_{NN}^{PW\rightarrow
PW'}(p_i,p)}{(2M_N)^2}\cdot \frac{M_N}{p^2-p_2^2-i\epsilon} \cdot \int \frac{l^2dl}{2\pi^2}~ \CB_
D^{PW'}(p)\nonumber \\
&& \cdot\frac{M_N}{p^2-p_1^2-i\varepsilon}\cdot
\CB_A(p) \ ,
\\
\CA_{ImSI}^{c,PW}&=&N_S\cdot  \int \frac{l^2dl}{2\pi^2}~\CB^{PW}_ D(p_i,l)
\cdot\frac{M_N}{l^2-p_1^2-i\epsilon}\cdot \int
\frac{k^2dk}{2\pi^2}~\frac{\CM^{^3P_1\rightarrow
^3P_1}_{NN}(l,k)}{(2M_N)^2} \nonumber \\ && \cdot
\frac{M_N}{k^2-p_1^2-i\epsilon}\cdot
\CB_A(k) \ , \\
\CA_{ISI+ImSI}^{c,PW}&=&N_S\cdot \sum_{PW'}^{^3S_1,^3D_1}\int
\frac{p^2dp}{2\pi^2}~\frac{\CM_{NN}^{PW\rightarrow
PW'}(p_i,p)}{(2M_N)^2}\cdot \frac{M_N}{p^2-p_2^2-i\epsilon}\cdot \int \frac{l^2dl}{2\pi^2}~\CB_
D^{PW'}(p,l)
\cdot\frac{M_N}{p^2-p_1^2-i\varepsilon}\nonumber
 \\ &&\cdot \int \frac{k^2dk}{2\pi^2}~\frac{\CM_{NN}^{^3P_1\rightarrow ^3P_1}(l,k)}{(2M_N)^2}
\cdot\frac{M_N}{l^2-p_1^2-i\epsilon}\cdot
\CB_A(k) \ . \Ee
The corresponding amplitudes of Figs.~\ref{feyn_fig1}d) can be obtained by changing
$A$ to $B$ in the above equations.

\subsection{Amplitudes of Fig.~\ref{feyn_fig1}e)-h)}
For the diagrams of Fig.~\ref{feyn_fig1}e)-h) there is no
intermediate state interaction so that the amplitude can be written as:
\Be \CA^{PW}= \CA_0^{PW} + \CA_{ISI}^{PW}. \Ee
Note that only the sum of the amplitudes of
Figs.~\ref{feyn_fig1}g)-i) is independent of the choice of the pion field. That is why
below we give only the amplitude for the  sum of them. The partial wave
amplitudes of Figs.~\ref{feyn_fig1}e)-i) without the ISI can be written as:

\Be \CA_0^{e,^3S_1}
&=&N_S^e\frac{4M_N^{\frac{3}{2}}m_\pi^2}{f_\pi^2}(-4c_1-2c_2+\frac{g_A^2}{4M_N}-2c_3) \cdot
u(p) \ ,
\\
\CA_0^{e,^3D_1} &=& -N_S^e\frac{4M_N^{\frac{3}{2}}m_\pi^2}{f_\pi^2}(-4c_1-2c_2+\frac{g_A^2}{4M_N}-2c_3)
\cdot w(p) \ , \Ee

\Be
\CA_0^{f,^3S_1} &=&N_S^f \frac{2m_\pi^2M_N^{\frac{3}{2}}}{f_\pi^4}\cdot
\int\frac{l^2dl}{2\pi^2}~u(l)
\cdot \int
\frac{d\Omega_{\hat{p}}}{4\pi}\frac{d\Omega_{\hat{l}}}{4\pi}
\frac{1}{(\vec p - \vec l~)^2+m_\pi^2} \ , \\
\CA_0^{f,^3D_1}
&=&N_S^f\frac{m_\pi^2M_N^{\frac{3}{2}}}{f_\pi^4}\cdot\int\frac{l^2dl}{2\pi^2}~w(l) \cdot \int
\frac{d\Omega_{\hat{p}}}{4\pi}\frac{d\Omega_{\hat{l}}}{4\pi}
\frac{1-3(\hat p \cdot \hat l)^2}{(\vec p - \vec
l~)^2+m_\pi^2} \ , \Ee

\Be \CA_0^{g+h+i,^3S_1}&=&N_S^{g+h+i}\frac{7\sqrt{2}g_A^2M_N^{\frac{3}{2}}m_\pi^2}{3f_\pi^4}\int\frac{l^2
dl}{2\pi^2}~
\int\frac{d\Omega_{\hat{p}}}{4\pi}\frac{d\Omega_{\hat{l}}}{4\pi}
\frac{1}{((\vec{p}-\vec{l})^2+m_\pi^2)^2}\nonumber \\ && \cdot \left [ \left (w(l)-\frac{u(l)}{\sqrt{2}} \right )\cdot (\vec{p}-\vec{l})^2-3w(l)\left (p(\hat{p}\cdot\hat{l})-l \right )^2 \right ] \ , \\
\CA_0^{g+h+i,^3D_1}&=&N_S^{g+h+i}\frac{7g_A^2 M_N^{\frac{3}{2}}m_\pi^2}{3f_\pi^4}\int\frac{l^2
dl}{2\pi^2}~
\int\frac{d\Omega_{\hat{p}}}{4\pi}\frac{d\Omega_{\hat{l}}}{4\pi}
\left [\frac{u(l)}{\sqrt{2}} \cdot (4p^2-8\vec{p}\cdot\vec{l}+6l^2(\hat{l}\cdot\hat{p})^2-2l^2) \right . \nonumber
 \\&& \left .  + \frac{w(l)}{2} \cdot \left (p^2-l^2+4\vec{p}\cdot\vec{l}-3p^2(\hat{p}\cdot\hat{l})^2-3l^2(\hat{p}\cdot\hat{l})^2\right ) \right ]\nonumber \\ && \cdot
\frac{1}{((\vec{p}-\vec{l})^2+m_\pi^2)^2} \ .
 \Ee

The corresponding amplitudes with the ISI can be obtained through
the following expression: \Be \CA_{ISI}^{m,PW}&=&N_s^m\cdot
\sum_{PW'}^{^3S_1,^3D_1}\int
\frac{p^2dp}{2\pi^2}~\frac{\CM_{NN}^{PW\rightarrow
PW'}(p_i,p)}{(2M_N)^2}\cdot \frac{M_N}{p^2-p_2^2-i\epsilon} \cdot \CA_
0^{m,~PW'}(p) \ , \Ee where $m$ can be 'e', 'f', or 'g+h+i'.
The symmetry factors for these amplitudes are
$N_s^e=\frac{8}{\sqrt{2}}$, $N_s^f=\frac{4}{\sqrt{2}}$ and $N_s^{g+h+i}=\frac{4}{\sqrt{2}}$.

\section{Observables}
In this appendix we present the expressions for the cross section
near threshold. The amplitudes of $pn\to d\pi\pi$ considered in this study
are calculated at the threshold and, therefore, can be
factored out of the phase space integration.
 The cross section for $pn\to d\pi^0\pi^0$ is expressed in terms of the amplitudes
 given in Appendix A in the following way: \Be
\sigma=\frac{1}{4\cdot 2\cdot 2}\cdot\left (
3|\CA^{^3S_1}_{full}|^2+3|\CA^{^3D_1}_{full}|^2 \right )\cdot
\frac{\Phi}{4E_{\tiny{CM}}\,p_{\tiny{CM}}} \label{xsection}\Ee where $E_{CM}$ and
$p_{CM}$ are the energy and momentum of the initial nucleons in the c.m.
frame. $\Phi$ is the phase space factor  defined as: \Be \Phi =
\int (2\pi)^4
\delta^4(P-p_{\pi_1}-p_{\pi_2}-p_d)\frac{d^3p_{\pi_1}}{(2\pi)^32E_{\pi_1}}\frac{d^3p_{\pi_2}}{(2\pi)^32E_{\pi_2}}
\frac{d^3p_{d}}{(2\pi)^32E_{d}} \Ee The prefactor $\frac{1}{4\cdot
2\cdot 2} $ is due to averaging over  the initial spin states, the isospin
wave function of the initial $NN$ states and the identity factor for the final pions,
respectively. The factor of 3 in Eq.~(\ref{xsection}) is
from summing up the spin states.


\begin{thebibliography}{99}




\bibitem{koltunundreitan}
  D.~S.~Koltun and A.~Reitan,
  Phys.\ Lett.\ {\bf 141} (1966) 1413.


\bibitem{report}
  C.~Hanhart,
  Phys.\ Rept.\  {\bf 397} (2004) 155 [arXiv:hep-ph/0311341].


\bibitem{heavymeson}  T.~S.~H.~Lee and D.~O.~Riska,
  Phys.\ Rev.\ Lett.\  {\bf 70} (1993) 2237;
 C.~J.~Horowitz, H.~O.~Meyer and D.~K.~Griegel,
  Phys.\ Rev.\  C {\bf 49} (1994) 1337
  [arXiv:nucl-th/9304004];
 J.~A.~Niskanen,
  Phys.\ Rev.\  C {\bf 53} (1996) 526
  [arXiv:nucl-th/9502015].

\bibitem{offshell}  E.~Hern\'andez and E.~Oset,
  Phys.\ Lett.\  B {\bf 350} (1995) 158
  [arXiv:nucl-th/9503019];
 C.~Hanhart, J.~Haidenbauer, A.~Reuber, C.~Sch\"utz and J.~Speth,
  Phys.\ Lett.\  B {\bf 358} (1995) 21
  [arXiv:nucl-th/9508005].


\bibitem{ourpols}
 C.~Hanhart, J.~Haidenbauer, O.~Krehl and J.~Speth,
  Phys.\ Rev.\  C {\bf 61} (2000) 064008 [arXiv:nucl-th/0002025].

\bibitem{Weinberg:1990rz}
  S.~Weinberg,
  Phys.\ Lett.\  B {\bf 251} (1990) 288.

\bibitem{Weinberg:1991um}
  S.~Weinberg,
  Nucl.\ Phys.\  B {\bf 363} (1991) 3.


\bibitem{NNpi}
B.Y. Park et al., Phys. Rev. C {\bf 53} (1996) 1519 [arXiv:nucl-th/9512023].



\bibitem{NNpicharged}
  C.~Hanhart, J.~Haidenbauer, M.~Hoffmann, U.-G.~Mei{\ss}ner and J.~Speth,
  Phys.\ Lett.\ B {\bf 424} (1998) 8 [arXiv:nucl-th/9707029].


\bibitem{BKMnovel}
  V.~Bernard, N.~Kaiser and U.-G.~Mei\ss ner,
  Eur.\ Phys.\ J.\  A {\bf 4} (1999) 259 [arXiv:nucl-th/9806013].


\bibitem{withandreas}
  C.~Hanhart and A.~Wirzba,
  Phys.\ Lett.\  B {\bf 650} (2007) 354 [arXiv:nucl-th/0703012];



\bibitem{NNpiloops}
 V. Dmitra\v sinovi\' c, K. Kubodera, F. Myhrer and  T. Sato,
{Phys. Lett.} B {\bf  465} (1999) 43 [arXiv:nucl-th/9902048];
  S.~I.~Ando, T.~S.~Park and D.~P.~Min,
  Phys.\ Lett.\ B {\bf 509} (2001) 253 [arXiv:nucl-th/0003004].

\bibitem{bira1} T.D. Cohen, J.L. Friar, G.A. Miller and U. van Kolck,
{Phys. Rev.} C {\bf 53} (1996) 2661 [arXiv:nucl-th/9512036].




\bibitem{rocha}
  C.~da Rocha, G.~Miller and U.~van Kolck,
  Phys.\ Rev.\ C {\bf 61} (2000) 034613 [arXiv:nucl-th/9904031].


\bibitem{ch3body}
 C. Hanhart, U. van Kolck, and
  G.A. Miller, Phys. Rev. Lett. {\bf 85} (2000) 2905 [arXiv:nucl-th/0004033].

\bibitem{withnorbert}
C.~Hanhart and N.~Kaiser,
Phys.\ Rev.\ C {\bf 66} (2002) 054005 [arXiv:nucl-th/0208050].


\bibitem{p-wave}
  V.~Baru, E.~Epelbaum, J.~Haidenbauer, C.~Hanhart, A.~E.~Kudryavtsev, V.~Lensky and U.~-G.~Mei{\ss}ner,
  Phys.\ Rev.\  C {\bf 80} (2009) 044003
  [arXiv:0907.3911 [nucl-th]].

\bibitem{lensky1}
  V.~Lensky, V.~Baru, J.~Haidenbauer, C.~Hanhart, A.~E.~Kudryavtsev and U.-G.~Mei{\ss}ner,
  Eur.\ Phys.\ J.\ A {\bf 27} (2006) 37 [arXiv:nucl-th/0511054].

\bibitem{Gardestig}
  A.~G{\aa}rdestig, D.~R.~Phillips and C.~Elster,
  Phys.\ Rev.\  C {\bf 73} (2006) 024002
  [arXiv:nucl-th/0511042].


\bibitem{Baru_proc}
  V.~Baru, J.~Haidenbauer, C.~Hanhart, A.~E.~Kudryavtsev, V.~Lensky and U.-G.~Mei{\ss}ner,
{\it Proceedings of the 11th International Conference
on Meson-Nucleon Physics and the Structure of the Nucleon (MENU 2007)},
J\"ulich, Germany, Sept. 10-14, 2007, p. 128 [arXiv:0711.2748 [nucl-th]].


\bibitem{kim}
 Y.~Kim, T.~Sato, F.~Myhrer and K.~Kubodera,
Phys.\ Rev.\  C {\bf 80} (2009) 015206
[arXiv:0810.2774 [nucl-th]].

\bibitem{jouni}
 J.~A.~Niskanen,
  Nucl.\ Phys.\  A {\bf 298} (1978) 417.


\bibitem{ourdelta}
 C.~Hanhart, J.~Haidenbauer, O.~Krehl and J.~Speth,
  Phys.\ Lett.\  B {\bf 444} (1998) 25 [arXiv:nucl-th/9808020].



\bibitem{Ars}
  A.~Filin, V.~Baru, E.~Epelbaum, J.~Haidenbauer, C.~Hanhart, A.~E.~Kudryavtsev and U.-G.~Mei{\ss}ner,
  Phys.\ Lett.\  B {\bf 681} (2009) 423 [arXiv:0907.4671 [nucl-th]].

\bibitem{Bolton}
  D.~R.~Bolton and G.~A.~Miller,
  Phys.\ Rev.\  C {\bf 81} (2010) 014001
  [arXiv:0907.0254 [nucl-th]].

\bibitem{KMN}
  U.~van Kolck, J.~A.~Niskanen and G.~A.~Miller,
  Phys.\ Lett.\  B {\bf 493} (2000) 65
  [arXiv:nucl-th/0006042].





\bibitem{ruso}L. Alvarez-Ruso, E. Oset and E. Hern\'andez, \npa{633} (1998) 519
[arXiv:nucl-th/9706046].

\bibitem{ruso2}
  L.~Alvarez-Ruso,
  Phys.\ Lett.\ B {\bf 452} (1999) 207
  [arXiv:nucl-th/9811058].

\bibitem{cao}
X.~Cao, B.~S.~Zou and H.~S.~Xu, arXiv:1004.0140 [nucl-th].



\bibitem{Kren}
  F.~Kren {\it et al.}  [CELSIUS/WASA Collaboration],
  Phys.\ Lett.\  B {\bf 684} (2010) 110
  [arXiv:0910.0995 [nucl-ex]].

\bibitem{Bashkanov}
  M.~Bashkanov {\it et al.},
  Phys.\ Rev.\ Lett.\  {\bf 102} (2009) 052301
  [arXiv:0806.4942 [nucl-ex]].


\bibitem{NNpipidata}
 T.~Skorodko {\it et al.},
  Phys.\ Lett.\  B {\bf 679} (2009) 30 [arXiv:0906.3087 [nucl-ex]].

\bibitem{Johanson}
  J.~Johanson {\it et al.},
  Nucl.\ Phys.\  A {\bf 712} (2002) 75, and references therein.



\bibitem{NNpipipol}
  S.~Abd El-Bary {\it et al.}  [COSY-TOF Collaboration],
  Eur.\ Phys.\ J.\  A {\bf 37} (2008) 267.
  [arXiv:0806.3870 [nucl-ex]].


\bibitem{schneider}
 S.~Schneider, S.~Krewald and U.-G.~Mei\ss ner,
  Eur.\ Phys.\ J.\  A {\bf 28} (2006) 107.

\bibitem{Kamano1}
  H.~Kamano, B.~Juli\'a-D\'{\i}az, T.~S.~Lee, A.~Matsuyama and T.~Sato,
  Phys.\ Rev.\  C {\bf 80} (2009) 065203
  [arXiv:0909.1129 [nucl-th]].

\bibitem{Kamano2}
  H.~Kamano, B.~Juli\'a-D\'{\i}az, T.~S.~Lee, A.~Matsuyama and T.~Sato,
  Phys.\ Rev.\  C {\bf 79} (2009) 025206
  [arXiv:0807.2273 [nucl-th]].


\bibitem{Gomez}
  J.~A.~G\'omez Tejedor and E.~Oset,
  Nucl.\ Phys.\  A {\bf 600} (1996) 413
  [arXiv:hep-ph/9506209].



\bibitem{BKMSgamma}
  V.~Bernard, N.~Kaiser, U.-G.~Mei{\ss}ner and A.~Schmidt,
  Nucl.\ Phys.\  A {\bf 580} (1994) 474
  [arXiv:nucl-th/9403013].

\bibitem{BKMgamma}
  V.~Bernard, N.~Kaiser and U.-G.~Mei{\ss}ner,
  Phys.\ Lett.\  B {\bf 382} (1996) 19
  [arXiv:nucl-th/9604010].



\bibitem{BKM_pipiN95}
  V.~Bernard, N.~Kaiser and U.-G.~Mei{\ss}ner,
  Nucl.\ Phys.\  B {\bf 457} (1995) 147
  [arXiv:hep-ph/9507418].



\bibitem{Fettes_pipiN}
  N.~Fettes, V.~Bernard and U.-G.~Mei{\ss}ner,
  Nucl.\ Phys.\  A {\bf 669} (2000) 269
  [arXiv:hep-ph/9907276].

\bibitem{Mobed}
  N.~Mobed, J.~Zhang and D.~Singh,
  Phys.\ Rev.\  C {\bf 72} (2005) 045204.


\bibitem{Kotulla}
  M.~Kotulla {\it et al.},
  Phys.\ Lett.\  B {\bf 578} (2004) 63
  [arXiv:nucl-ex/0310031].


\bibitem{Chiral_L_NN1}
C. Ord\'o\~nez and U. van Kolck, Phys. Lett. B {\bf 291} (1992) 459;
U. van Kolck, U.T. Ph.D. (1993); C. Ord\'o\~nez, L. Ray, and U. van
Kolck, Phys. Rev. Lett. {\bf 72} (1994)
 1982; Phys. Rev. C {\bf 53} (1996) 2086.


\bibitem{Chiral_L_NN2}
 V.~Bernard, N.~Kaiser and U.-G.~Mei{\ss}ner,
  Int.\ J.\ Mod.\ Phys.\  E {\bf 4} (1995) 193 [arXiv:hep-ph/9501384].

\bibitem{bernard_rev}
  V.~Bernard,
  Prog.\ Part.\ Nucl.\ Phys.\  {\bf 60} (2008) 82
  [arXiv:0706.0312 [hep-ph]].



\bibitem{NN1}
P.~F. Bedaque and U.~van Kolck,
Ann. Rev. Nucl. Part. Sci. {\bf 52} (2002) 339
 [arXiv:nucl-th/0203055].
\bibitem{NN2}
E.~Epelbaum,
Prog. Part. Nucl. Phys. {\bf 57} (2006) 654 [arXiv:nucl-th/0509032].
\bibitem{NN3}
  E.~Epelbaum, H.-W.~Hammer and U.-G.~Mei{\ss}ner,
  Rev.\ Mod.\ Phys.\  {\bf 81} (2009) 1773
  [arXiv:0811.1338 [nucl-th]].





\bibitem{ccf}
  J.~Haidenbauer, K.~Holinde and M.~B.~Johnson,
  Phys.\ Rev.\  C {\bf 48} (1993) 2190.

\bibitem{cdbonn} R.~Machleidt, \prc{63} (2001) 024001.

\bibitem{cdbonn_delta}
A.~Deltuva, R.~Machleidt, and P.~U.~Sauer, \prc{68} (2003) 024005;
A.~Deltuva, private communication.


\bibitem{epelbaum}
H.~ Krebs, E. ~Epelbaum and U.-G. Mei{\ss}ner, \epja{32} (2007) 127.

\bibitem{bernard}
V.~Bernard, N.~Kaiser and U.-G. Mei{\ss}ner, \npa{615} (1997) 483.

\bibitem{pid_isospinviol}
  V.~Baru, C.~Hanhart, M.~Hoferichter, B.~Kubis, A.~Nogga and D.~R.~Phillips,
  arXiv:1003.4444 [nucl-th].



\bibitem{lensky_thesis}
V. Lensky, PhD Thesis, University of Bonn (2007).

\bibitem{dNN}
V. Tarasov, V. Baru and A. Kudryavtsev, Phys. At. Nucl. {\bf 63} (2000) 801.

\end{thebibliography}
\end{document}